# A Survey on Cross-Architectural IoT Malware Threat Hunting

**ANANDHARAJU DURAI RAJU**[ID]1, **IBRAHIM Y. ABUALHAOL**[ID]2, **(Senior Member, IEEE),**
**RONNIE SALVADOR GIAGONE**[ID]3, **YANG ZHOU**[ID]4, **AND SHENGQIANG HUANG**[ID]3

1School of Computing Science, Simon Fraser University, Burnaby, BC V5A 1S6, Canada
2Huawei Technologies Canada Company Ltd., Kanata, ON K2K 3J1, Canada
3Huawei Technologies Canada Company Ltd., Burnaby, BC V5C 6S7, Canada
4Huawei Technologies Canada Company Ltd., Markham, ON L3R 5A4, Canada

Corresponding author: Anandharaju Durai Raju (anandharaju@ieee.org)

**ABSTRACT** In recent years, the increase in non-Windows malware threats had turned the focus of the cybersecurity community. Research works on hunting Windows PE-based malwares are maturing, whereas the developments on Linux malware threat hunting are relatively scarce. With the advent of the Internet of Things (IoT) era, smart devices that are getting integrated into human life have become a hackers' highway for their malicious activities. The IoT devices employ various Unix-based architectures that follow ELF (Executable and Linkable Format) as their standard binary file specification. This study aims at providing a comprehensive survey on the latest developments in cross-architectural IoT malware detection and classification approaches. Aided by a modern taxonomy, we discuss the feature representations, feature extraction techniques, and machine learning models employed in the surveyed works. We further provide more insights on the practical challenges involved in cross-architectural IoT malware threat hunting and discuss various avenues to instill potential future research.

## I. INTRODUCTION

Each day the digital world is exposed to millions of new malware (Malicious Software) attacks, and unfortunately, almost all of them are oblivious to the day-to-day users while they happen. In the past two decades, the machine learning approaches adapted to the domain of malware detection/classification strove towards convergence at better handling of malware threats as hard as zero-day attacks. In recent years, deep learning approaches are also taking part in the arena to cope with the explosion of malware variants.

Despite the COVID-19 pandemic, malware attacks are steadily on the rise [1], [2]. While most financially motivated attack groups focus on big industry players, surprisingly, more than 60% of the attacks are directed towards small and mid-sized businesses [3]. Such cybercrimes are estimated to cause a world-wide damage of 6 Trillion in 2021 [3] and are expected to reach 10.5 Trillion by 2025. Particularly, ransomware-based attacks grew by more than 150% since 2020, with extortion demands exceeding millions in the top cases [4].

Several malware groups such as Egregor, and Netwalker, have started providing Ransomware-as-a-Service (RaaS), accounting for up to 64% of the total ransomware attacks in 2020. Recently, Sierra Wireless, an IoT (Internet of Things) product vendor, suffered a ransomware attack, leading to disruption of its IT operations and production halt [5].

The IoT devices are deemed to be the most targeted at present times. Even the wearables like FitBit device's applications are also vulnerable to getting hacked and may allow a threat actor to tap onto a wealth of PII (Personally Identifiable Information) of end-users [6]. Smartly connected devices such as security cameras, refrigerators, and toasters were surprisingly part of the BotNets (roBot Networks) in the infamous massive DDoS (Distributed Denial of Service) cyber-attack against Dyn DNS provider by Mirai, which caused parts of the world inaccessible to major sites like Airbnb, Twitter, PayPal, GitHub, Amazon, Netflix [7]. Following the Mirai attack, DDoS attacks via BotNets are now the extensively used distributed attack source targeting

IoT devices, and their strains spread over 25 different malware families [8]. DDoS-for-hire had become one of the trending hack-for-hire services, where botnets with GBps to TBps attack bandwidth are being sold in the underground forums of the dark web. In early 2019, honeypots tracked and monitored by Kaspersky Labs found about 105 million such IoT device DDoS attacks that originated from 276,000 unique IPs.

In light of the above issues, the problem of malware detection and/or classification continues to be a topic of much importance. We address this problem of detecting and/or classifying the malware threats commonly with the term '*Malware Threat Hunting*'.

While there are a vast number of works on Windows PE (Portable Executable) based malwares such as [9]–[12], [13]–[16], [17]–[19], the research works proposed for IoT-based malwares are comparatively scarce especially in the context of cross-architectural IoT malware threat hunting. This is due to the complexity that naturally arises when there are multiple CPU architectures, OS platforms, and diverse target devices to be taken care of by a single learning approach [20], [21].

There are few valuable survey works that exist on IoT malwares, such as [22] for static analysis and [23] for dynamic analysis. The purpose of this survey is to complement the existing works as well as provide insights into the critical cross-architectural dimension covering the latest developments.

The main contributions of this survey are: (1) to bridge the gaps in existing surveys by providing a comprehensive review on the recent developments in research methods for IoT malware threat hunting, (2) to discuss the benefits and limitations of those recent works, (3) to provide a modern taxonomy of features for learning malwares and analyzing their usability for detecting cross-architectural threats, and (4) to explore the challenges and issues in conducting research in the context of cross-architecture IoT malware threat hunting. Furthermore, (5) it highlights the gap in the existing literature and discusses various directions for potential future research.

The rest of this article is organized as follows: Section II provides a brief background on IoT-based malware and a brief overview of the ELF (Executable and Linkable Format) specification and useful tools for feature extraction. Section III describes the modern taxonomy of malware detection approaches based on the type of features used, and Section IV systematically discusses the surveyed papers along with the feature representations used and feature selection mechanisms employed. The existing challenges and future research opportunities are discussed in Section V. Section VI discusses the related survey works in terms of general IoT ELF approaches and cross-architectural approaches. Section VII concludes the study.

## II. IoT MALWARE BACKGROUND
Internet-of-Things (IoT) is a large set of devices connected via the private or public internet, and that is infused with the ability to talk to each other streaming real-time data with less or no intervention required from humans, thereby building a unified intelligence.

Nowadays, devices of any size with a chip installed for enabling centralized control, device-to-device control, wireless sensor networks, and embedded systems are considered IoT devices. For example, security motion sensors, smartphones, voice assistant-controlled home automation devices like TVs, speakers, home lighting systems are considered IoT devices.

IoT devices are generally equipped with less computing and storage compared to traditional laptops and PCs. These reduced configurations impose tight constraints leading to the need for specialized operating systems (OS) and CPU architectures. While the OSes such as Windows, Linux, Android, and iOS dominate the PCs, Laptops, servers, and mobile devices, they are not directly suitable for embedded devices in the constrained IoT space, where deployments to a scale of millions of devices are expected in addition to satisfying the lower cost and economical operational constraints.

Therefore, an OS for IoT device should be lightweight to support the minimal hardware, yet following the security requirements, as suggested by Pal *et al.* [26], which includes energy awareness, security by privacy, location privacy, storage, communication, scalability by incremental deployment, load balancing, robustness, fault tolerance, connectivity, and usability. Linux flavors and distributions such as Ubuntu Core, Raspbian (Debian) support such requirements and hence are widely used by IoT developers.

Currently, the world of IoT devices is still poorly secured, inviting the threat actors with open doors. ELF-based malware gained attention from the cybersecurity community only in the mid of the past decade when a large number of samples started accumulating with VirusTotal [27], [28], before which it was generally believed that Linux was not as vulnerable as Windows. Threat hunting techniques proposed specifically to OSes such as Windows and Android would follow a one-of-a-kind approach and usually make use of features tailored for that specific OS and would not be transferable to other OSes. Techniques proposed for Linux ELF threat hunting suffer from not being able to follow the many-of-the-same-kind approach to accommodate the multiple distributions and variants within the Linux landscape.

Most known IoT malware indeed can be labeled more generically as Linux malware. Bots are malicious software programs that are designed to perform predetermined tasks using the resources of the host system that they infect. They could aid the attacker to compromise other devices in the host network [29] and steal confidential and/or sensitive information.

Figure 1 shows that Linux variants are the most utilized operating systems for IoT devices, according to a survey by Eclipse Foundation, 2018 [24]. As of 2020, Linux and FreeRTOS were the top OSes IoT developers preferred with 43% and 35% ratings, respectively. Figure 2 illustrates the ranking among distros within Linux.

**TABLE 1.** Types of segment entries in ELF [25].

| Segment Type | Description |
|---|---|
| PT_NULL | PT_NULL allows program header table to contain entries that can be ignored during execution |
| PT_LOAD | PT_LOAD is used to define a loadable segment using the values specified by p_memsz & p_filesz. Such loadable segment entries are sorted by by their p_vaddr member and are listed in the ascending order in the program header table |
| PT_DYNAMIC | PT_DYNAMIC is used to describe dynamic linking information |
| PT_INTERP | The segment type PT_INTERP meaningful only for executable files and possibly shared objects occurs once, preceding any of the loadable segment entry. It is used to describe the size and location of an interpreter to be invoked via a null terminated path |
| PT_NOTE | PT_NOTE segment is used to describe the size and location of auxiliary information |
| PT_SHLIB | Having unspecified semantics, the PT_SHLIB is a reserved segment type entry whose presence indicate that the binary may not conform to the ABI (Application Binary Interface) |
| PT_PHDR | The segment type PT_PHDR occurs once in a binary file preceding any of the loadable segment entry when program header table is part of the program's memory image, and is used to describe size and location of the program Header table in both the disk file as well as program's memory image |
| PT_TLS | PT_TLS is not a mandatory entry program header table. It is used for the specification of the template for thread-local storage. |
| PT_LOOS to PT_HIOS | Operating system specific semantics are described via these reserved entries |
| PT_LOPROC to PT_HIPROC | Processor-specific semantics are decribed via these reserved entries |

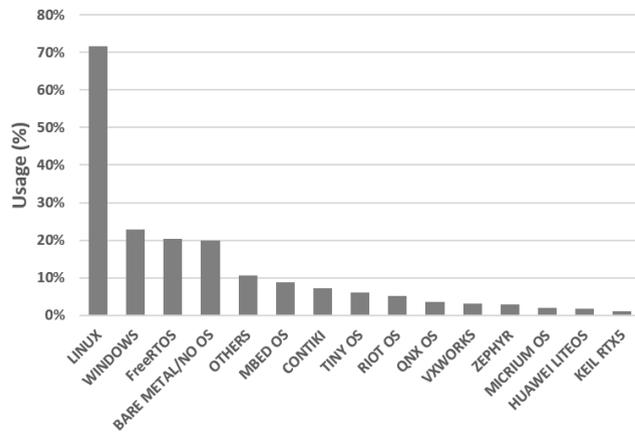

**FIGURE 1.** Ranking of operating systems for IoT [24].

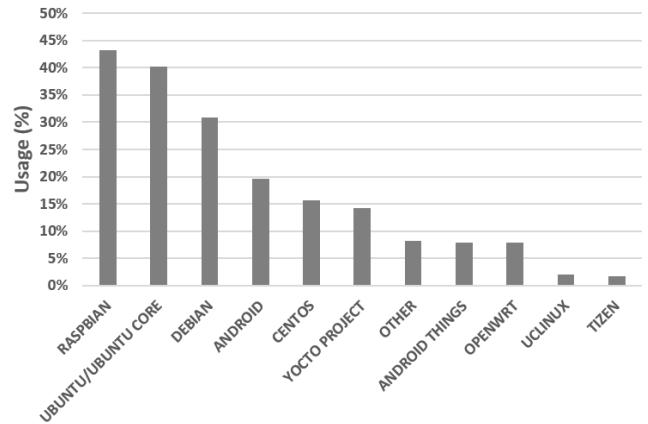

**FIGURE 2.** Ranking among linux distros for IoT [24] .

Lastly, much of the existing research work on IoT malware threat hunting declare the problem of handling the different CPU architectures such as MIPS, AARCH, and ARM, as the prominent challenge being encountered. This malware threat hunting problem is much more complex when the wide and diverse environments of the target devices are considered, such as medical equipment, wearables, scanners, and security cameras. Linux program binaries need not necessarily specify the target environment for which they are configured to run. Such a lack of standards further amplifies the problem of threat hunting.

### A. ELF FILE FORMAT

Executable and Linkable Format (*ELF*) is the standard binary file format [45] for the file types Linux executables, shared libraries, core dumps, and object codes, used by operating systems like Linux, BSD, Solaris, BeOS, and Android. It was initially specified for Unix System V by Unix System Laboratories (USL) and Unix International (UI). The cross-platform properties of ELF allowed it to be used across different CPU architectures: Intel (x86, x64), ARM, MIPS, Motorola, SPARC, PowerPC, Renesas SH, Motorola m68k, and different target devices: Routers, Printers, Cameras, Smart TVs, medical devices, and VoIP. The ELF file format also allows Linux programs to specify arbitrary loaders.

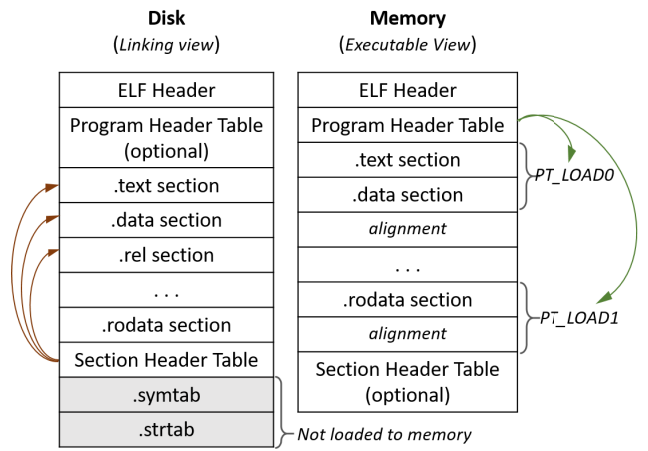

**FIGURE 3.** ELF file format - source [44].

**TABLE 2.** Tools for ELF static analysis.

| Tool | Used By | Purpose of Use / Description |
|------|---------|------------------------------|
| bindiff | [21] | Binary file analysis tool for disassembled code similarity and function similarity |
| binwalk | [30] | Extracting firmware images and reverse engineering |
| diaphora | [23] | Advanced binary program diffing tool with support to assembler and CFG diffing, call graph matching calculation, etc |
| elfdump | - | Available under Solaris and FreeBSD. Useful to find detailed information about dynamic linkages, relocations, non-stripped binary's symbol details, dependencies on shared objects, functions, sections and program segments [31] |
| elfutils | [32] | Faster, more featureful alternative tools to GNU Binutils purely for Linux [33] |
| file | [8], [30] | Useful to determine type of a file - not to be used as a security tool as it can be easily fooled by abusing a file's magic |
| ghidra | [34] | Reverse engineering tool like IDA. Extensible, supports analysis of very large firmware images, ability to decompile object code back to source code |
| hexdump | - | Utility for inspecting files via hex, decimal, octal and ASCII views. Allows data recovery and reverse engineering |
| hexedit | - | Helps to view/edit files in hex or ASCII |
| IDAPro | [28], [21], [34], [35], [36], [30] | Prominently used interactive disassembler and debugger tool |
| magic | - | file command's magic pattern file |
| nucleus | [28] | A structural control flow graph analysis based compiler agnostic function detection tool for binaries proposed by Andriesse *et al.* [37]. |
| obj(ect)dump | [38], [39], [40] | Information dump about object files including intended target instruction set architecture (ISA) and structural information. Relies on BFD. |
| od (octal dump) | - | Tool for debugging, visualizing executable code, and dumping in octal (default), hex, ASCII formats. |
| openwrt | [21] | For benign firmwares |
| pyelftools | [28] | Python library to parse and analyze ELFs and debugging |
| radare2 | [32], [34], [41] | Binary forensic analysis, reverse engineering, exploiting and debugging tool. Options such as 'afl' can be used to disassemble function lists, get count of functions etc. |
| readelf | [28], [42], [8], [43] | Prominently used to obtain ELF structural information. Provides more details than objdump. It is independent of Binary File Descriptor (BFD) library. |
| size | [8] | Provides total ELF size as well as section sizes |
| strings | [8], [1] | ASCII strings information from binary |

Figure 3 illustrates the general ELF file format [44]. ELF file is composed of three major categories: Program Header that aids in handling memory segments during run time execution by providing information to the system on how to create process images, the individual 'sections' that hold various types of information such as 'code' and 'text,' and finally, the Section Header that describes the various file sections such as their offset information and also helps in linking and relocation process.

Shared objects and executables employ the program header that provides an array of structural information describing segments and other information needed to execute a program. Each segment can represent one or more sections, and a binary must contain at least one loadable segment to allow the system to load it. However, this requirement is not mandated by the file format [25]. Table 1 provides a list of some segment types typically found in an ELF binary.

There are two types of views: 1) *Linking view*, where the sections and the section header table are important but the program header table is optional; 2) *Execution view*, where the segments and the program header table are important, but the section header table is optional.

Windows PE file format and Linux ELF file format are similar in structure in that both use a Header that defines meta-information about the rest of the file structure, and in that, both formats use a Section Header to define the individual sections. Unlike PE format, ELF additionally uses the 'Program Header' table that steps into action during runtime.

The ELF holds other essential data for runtime execution support, to aid debugging, and to provide a human-readable view, such as symbols representing the symbolic names to functions and data [42]. There are two different symbol table sections found in the shared objects and dynamic executables: 1) '.symtab' that keeps information to locate or relocate symbolic definitions and references of a program, and 2) 'dynsym' - a smaller version of the symtab containing global data. The dynsym table also helps to understand the runtime functionality and the expected behaviors of an ELF program by determining the system calls that the executable could import during run time. The system call address that this table provides can be used for further debugging.

### B. IoT CPU ARCHITECTURES

The rapid proliferation of IoT devices that can perform an assortment of functionalities calls for complex product design across the IoT landscape to achieve high performance with low power demands. For instance, ARM follows a reduced instruction set architecture (RISC) that requires less hardware and less power than a complex instruction set architecture (CISC) such as x86. This makes ARM more suitable for wearable technologies, while x86 chips are more suitable for laptops and desktops [46].

Each CPU architecture in the IoT market is built for a specific purpose under various constraints that arise due to the trade-offs between power and performance. Such constraints include memory address width and hierarchy, depth of processor pipeline, size of the data bus, operation frequencies, and out-of-order execution. The complexity is compounded with recent developments in IoT to support Artificial Intelligence and Machine learning tasks, which require far greater performance, power, and latency requirements. For instance, home-based IoT devices require motion tracking, speech recognition, and response and image analysis. These constraints and complexities give rise to different instruction set architectures and explain the need for the neutrality of selected feature representation across various instruction set architectures for effective malware threat hunting.

Some examples of different CPU architectures are x86, ARM, MIPS, SPARC, AARCH64, PowerPC, Renesas SH, Motorola 68020. More information on the instruction sets, opcodes, and assembly language specifications for various architectures can be found in these resources: [47]–[50], [51]–[53].

### C. IoT OS PLATFORMS
Similar to standard operating systems like Windows, iOS, and Linux, the IoT operating systems are expected to manage the embedded device functions but operate under the limited memory footprint, power, and processing capabilities, yet possess the properties such as portability, scalability, security, connectivity, modularity. Some open-source operating systems for IoT include Raspbian, Contiki, FreeRTOS, Ubuntu Core, ARM mbed, Yocto, Apache Mynewt, and Zephyr OS and some of the commercial IoT OSes include Windows 10 IoT, Android Things, WindRiver VxWorks, Freescale MQX, Mentor Graphics Nucleus RTOS, Express Logic ThreadX, TI RTOS and Particle. In light of these diverse OSes, it is crucial to choose feature representations with capabilities for OS platform independence.

### D. FEATURE EXTRACTION TOOLS FOR ELF STATIC ANALYSIS
A Linux-based operating system interprets the desired machine instructions using the formal ELF file format specification, which is the binary output format of a compiler or linker [54]. In Table 2, we provide a short overview of the tools, including the tools used in surveyed works, that are helpful to analyze, debug and extract useful information from ELF files. Many of the surveyed works used the scanning services such as VirusTotal, Shodan, and Zmap to label their ground truth and validate the datasets employed in their studies.

### E. MALWARE THREAT HUNTING APPROACHES
The malware analysis phases involved in the malware threat hunting process can be generally classified into static, dynamic, and hybrid analysis categories.

Static malware analysis where a binary file is reverse engineered, disassembled, or dissected using tools discussed in section 2 to analyze various structural and semantical

```
ELF 64-bit LSB shared object,
x86-64, version 1 (SYSV),
dynamically linked,
BuildID[sha1]=3b41c6707ba33ecc7afe...,
stripped
```

**FIGURE 4.** Basic information provided by 'file' command.

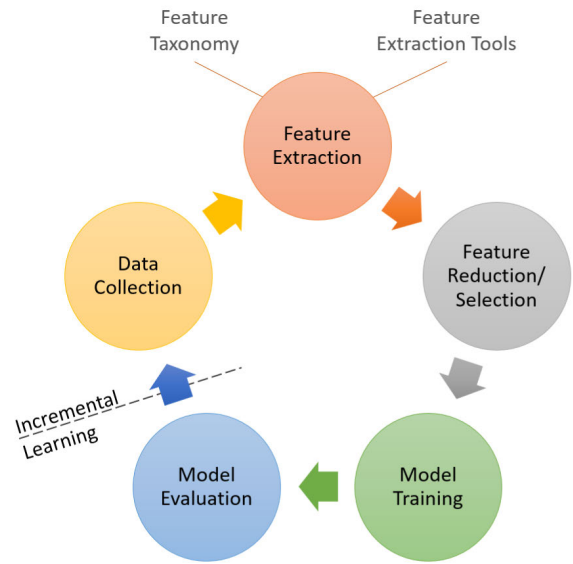

**FIGURE 5.** ML pipeline for static malware threat hunting process.

information found in the binary file without any execution. This method is susceptible to evasive methods like anti-disassembly, code obfuscation, and adversarial techniques. Dynamic analysis is a behavioral method that observes or debugs a malware's behavior in an isolated host such as sandboxes. Dynamic methods are also susceptible to evasive techniques such as anti-debugging and deferred execution. Hybrid analysis methods combine both static and dynamic methods and leverage their advantages.

Similar to Windows portable executables (PE), ELF binaries can also be broadly categorized into static and dynamically linked. Statically linked binaries include a copy of all libraries required for execution, making them more portable but bigger in size. Dynamically linked binaries depend on the execution environment to supply libraries needed during runtime.

Ngo et al. [22] claimed that the static analysis method has more ability than dynamic methods in analyzing malware structure without the need to consider processor architecture. Figure 4 provides a sample basic information that can be obtained using Linux 'file' command.

As shown in Figure 4, symbol, debugging, and relocation information could be stripped from an ELF binary to make them lightweight. However, studies have shown that IoT

**Taxonomy of ELF Feature Representations**

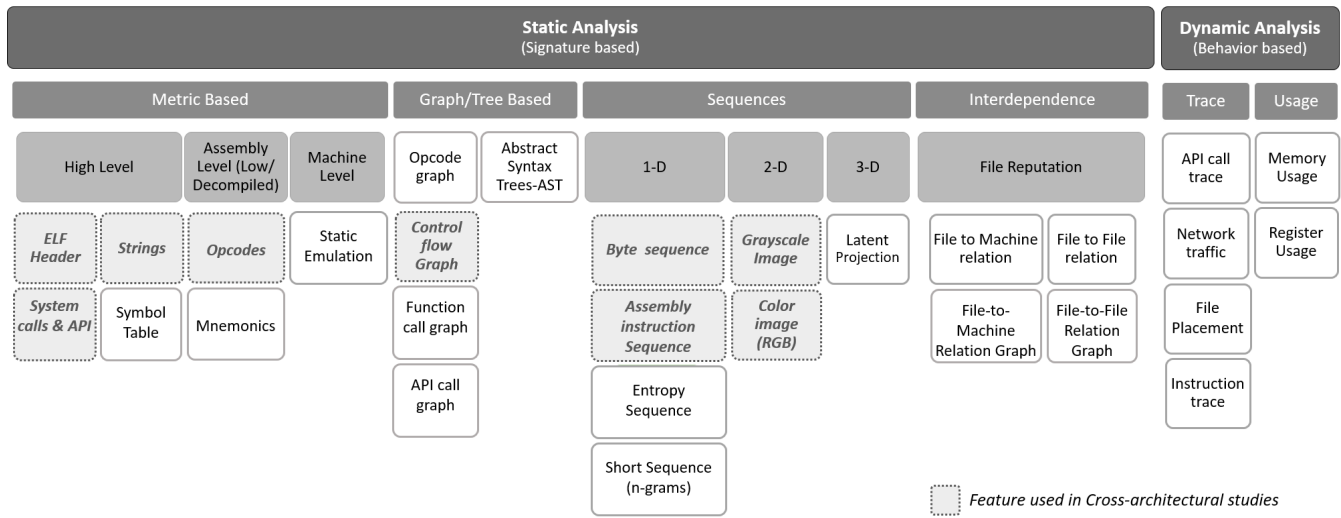

**FIGURE 6.** Taxonomy of ELF feature representations.

malwares are mostly statically linked [35] and not stripped to reduce the dependency on the diverse IoT execution environments and avoid runtime failures. It also makes them hard to analyze under static analysis.

Figure 5 illustrates the generic machine learning-based pipeline for static malware threat hunting. It also showcases where the feature extraction tools described in Section II-D and the modern taxonomy described in Section III fit the pipeline.

## III. TAXONOMY

In this section, we provide a taxonomy of feature representations used for static analysis-based malware threat hunting in the IoT landscape and highlight the features specifically useful for cross-architectural IoT malware threat hunting that requires the abilities of ISA neutrality and OS platform independence as described in Sections II-B and II-C respectively. The basis for this modern taxonomy is to account for the latest developments for such latter tasks and also provide the high-level gaps in existing research (see Section V-B).

Figure 6 provides the categorization of the feature representations based on four major divisions, namely: metric-based, graph/tree-based, sequence-based, and interdependence. These divisions encompass representations extracted from the content within a sample, such as strings and opcodes, as well as the external characteristics of a sample such as file-to-machine relations. We next provide a brief summary of these divisions and the feature types categorized under them.

### A. METRIC BASED
#### 1) HIGH-LEVEL FEATURES
This section describes the high-level informative metrics that can be extracted from binaries and used as features.

#### a: ELF HEADER
The ELF is the standard file specification/framework for executables, shared objects, and relocatables in a Linux-based system. ELF Header stores rich structural information that is important to support the framework [45], [54], such as file's magic data, class (32-bit or 64-bit), entry points, target application binary interface (ABI), file interpretation indicators, sizes, and addresses to program headers and section headers. ELF Header contains both machine-dependent and machine-independent features. Structural information from ELF headers was used by Alhanahnah *et al.* [21] and Shahzad and Farooq [43].

#### b: STRINGS
Concerning static malware analysis, a binary file may not be human-readable in its entirety. However, it may contain some human-readable strings or sequence of characters within the binary content [55], such as IP addresses, DLL names, error messages, and code comments.

#### c: SYMBOL TABLE
It acts as the lookup table holding the location and relocation information of symbolic references in a binary file to support the processes of linking and debugging. For instance, the non-allocable symbol table '.symtab' holds information on register symbols, local symbols, and section symbols that are not used during runtime. The allocable dynamic symbol table '.dynsym' holds global symbols that may be used during runtime execution [56]. Note that the '.symtab' can be stripped from a binary file to make it lightweight, while '.dynsym' is not.

#### d: SYSTEM CALLS AND APIs
System calls act as an interface to access the operating system provided services such as file and device management operations, controlling processes and communications, and maintenance of information, while application programming interfaces (APIs) are system call wrappers written in high-level languages. For instance, UNIX-based systems use POSIX APIs. Both system calls and APIs may also occur as printable strings in a binary.

**TABLE 3.** Architecture dependency of Mnemonics for *dvrHelper* (Mirai).

| Architecture | Mnemonic | |
|---|---|---|
| | Function call | Move operation |
| x86 | CALL | PUSH |
| ARM | BL | LDR |
| MIPS | JALR | LW |
| PPC | BL | LI |
| SPARC | CALL | MOV |

### 2) ASSEMBLY LEVEL FEATURES
#### a: OPCODES AND MNEMONICS
Opcodes (Operation codes) are unique and atomic executable instructions close to machine code. They accept registers and/or operands information as parameters to perform their intended operation. For instance, a move operation in the PowerPC instruction set is given by ''LI R4, 0'' comprises the mnemonic for move operation 'LI' (opcode = 38), the register R1, and the operand 0. Opcodes have proved to be more useful in detecting and classifying malwares [32], [38], [39]. Mnemonics are a special form of opcodes with symbolic names that are self-explanatory and easily understood by humans. However, both opcodes and mnemonics are generally dependent on the instruction set architecture. Tables 3 and 4 showcases such an example of architecture dependency using the Mirai botnet disassembly for its 'dvrHelper' function call - different notations are used for the same operation by different processor architectures. dvrHelper is a DDoS attack module equipped with features to bypass anti-DDoS solutions.

### 3) MACHINE LEVEL FEATURES
#### a: STATIC EMULATION
The static emulation is inspired by dynamic analysis on emulated environments using software tools like QEMU. Static emulation refers to the analysis of loadable parts of the program. For instance, in Figure 3, the segments $PT\_LOAD0$ and $PT\_LOAD1$ denote sections that will be loaded for execution during runtime. There may be other sections that will never be loaded in addition to '.symtab,' '.strtab.' Excluding such information and focusing on loadable parts is still a major gap in the existing studies.

### B. GRAPH/TREE-BASED FEATURES
### 1) GRAPH-BASED FEATURES
They are an extended version of metric-based features discussed above, where the relationship among the features is also accounted for and expressed. The nodes in the graph usually represent the actual metric-based features like APIs,

Opcodes, and their relationships are expressed by the edges. Several works have been proposed over Graph-based methods, such as opcode graphs by [57], control flow graphs by [41], function call graph by [30], PSI-graph by [30], and API-call graphs [58]. In general, graph-based methods suffer from the feature extraction overhead with more time spent on generating the graphs. There are also other graph-based methods proposed based on interdependence properties for analyzing file reputations, such as file-to-machine relation graphs by Chau et al. [59], Ye et al. [60], and file-to-file relation graphs with significant performance over the task of malware detection by Tamersoy et al. in [61]. Graph-based methods like CFG may provide potential information about the existence of packing and obfuscation [41].

### 2) TREE-BASED FEATURES
Abstract Syntax Trees (ASTs) are the tree representations generated using parsers over the code constructs found in a source code's syntactic structure, and tree-based machine learning approaches are later employed to learn the latent information they hold. Being a byproduct of the compiler's syntax analysis phase, ASTs are useful for analyzing or transforming programs to a more simplified view for better understanding.

Phan and Le Nguyen [62] employed deep neural networks on sequences of assembly instructions instead of Abstract Syntax Trees (AST) to improve feature-based and tree-based approaches for software defect prediction. The study discusses that ASTs only reflect the structures and do not reveal the behavior of programs, but assembly information does. The reason behind it is that ASTs tend to provide the structure of a program rather than internal behaviors [62]. However, ASTs have been successfully used to detect Powershell-based malware in [63] and Javascript-based malware in [64].

### C. SEQUENCE-BASED FEATURES
### 1) 1-D SEQUENCE
#### a: BYTE SEQUENCE
It is a sequential representation of byte-level data present in binary files, where each byte is converted into an 8-bit integer (unsigned) and translated to numerical representation with values ranging from 0 to 255. The encoding of text information in the actual binary is often overlooked, and a default encoding of UTF-8 is assumed. Several models work well under this assumption for Windows [65]–[67], and Linux [40] platforms, yet the effect of considering the actual encoding is largely unexplored.

#### b: ASSEMBLY INSTRUCTION SEQUENCE
Assembly instructions extracted from a disassembled binary are concatenated into a one-dimensional sequence. The operands and registers may be pruned out to reduce sequence length. Tokenization or embedding of the resulting sequence may be required.

**TABLE 4.** Disassembly showing 'dvrHelper' (Mirai) botnet call for different IoT processor architectures.

| | | | |
|---|---|---|---|
| **x86** | 0x080481f8 | 83c41c | add esp, 0x1c |
| | 0x080481fb | 68ff010000 | push 0x1ff ; 511 |
| | 0x08048200 | 6841020000 | push 0x241 ; 577 |
| | 0x08048205 | 683d840408 | push str.dvrHelper ; 0x804843d ; "dvrHelper" ; int32_t arg_8h |
| | 0x0804820a | 8945e4 | mov dword [var_1ch], eax |
| | 0x0804820d | e8f5feffff | call fcn.08048107 |
| **ARM** | 0x000081d8 | 20119fe5 | ldr r1, [0x00008300] ;[0x8300:4]=0x241 ; int32_t arg2 |
| | 0x000081dc | 84008de5 | str r0, [var_84h] |
| | 0x000081e0 | 1c219fe5 | ldr r2, [0x00008304] ;[0x8304:4]=511 |
| | 0x000081e4 | 1c019fe5 | ldr r0, [str.dvrHelper] ;[0x8368:4]=0x48727664 ;"dvrHelper";int32_t arg1 |
| | 0x000081e8 | b8ffffeb | bl fcn.000080d0 |
| **MIPS** | 0x004002a4 | a7a2001e | sh v0, 0x1e(sp) |
| | 0x004002a8 | 8fbc0010 | lw gp, 0x10(sp) |
| | 0x004002ac | 24050301 | addiu a1, zero, 0x301 |
| | 0x004002b0 | 8f848018 | lw a0, -segment.LOAD0(gp) ; [0x440638:4]=0x400000 segment.ehdr |
| | 0x004002b4 | 8f998060 | lw t9, -0x7fa0(gp) ; [0x440680:4]=0x400100 |
| | 0x004002b8 | 248405f0 | addiu a0, a0, 0x5f0 |
| | 0x004002bc | 240601ff | addiu a2, zero, 0x1ff |
| | 0x004002c0 | 0320f809 | jalr t9 |
| **PPC** | 0x10000274 | 90610010 | stw r3, 0x10(r1) |
| | 0x10000278 | 3c601000 | lis r3, 0x1000 |
| | 0x1000027c | 38800241 | li r4, 0x241 ; int32_t arg2 |
| | 0x10000280 | 38a001ff | li r5, 0x1ff ; int32_t arg3 |
| | 0x10000284 | 38630548 | addi r3, r3, 0x548 ; int32_t arg1 |
| | 0x10000288 | 4bfffe61 | bl fcn.100000e8 |
| **SPARC** | 0x00010204 | 9010204d | mov 0x4d, o0 |
| | 0x00010208 | 92102601 | mov 0x601, o1 |
| | 0x0001020c | d027bfe8 | st o0, [fp+-0x18] |
| | 0x00010210 | 941021ff | mov 0x1ff, o2 |
| | 0x00010214 | 11000040 | sethi 0x40, o0 |
| | 0x00010218 | 7fffffb5 | call fcn.000100ec |

### c: ENTROPY SEQUENCE

It is the sequence of rolling entropy obtained by scanning a series of short windows of byte sequences [28], assembly instruction sequences, or simply the whole file [8]. Various entropy measures proposed in the literature had been used for malware threat hunting scenario, such as Shannon entropy used in [40].

### d: SHORT SEQUENCE

This is a special case of very short byte sequences. They divide long sequences into several disjoint or overlapping short sequences, typically comprising sequences of 2 to 11-byte length, generally called n-gram byte sequences, where 'n' denotes the sequence length. Short 'n' value leads to too much granularity and faces a severe curse of dimensionality, while larger 'n' value leads to better accounting of long-range contextual relationships but loses granularity. N-grams are employed in studies such as [21].

### 2) 2-D SEQUENCE

### a: GRAYSCALE IMAGE

A two-dimensional image-like representation is obtained by reshaping and then resizing the one-dimensional byte sequence representation discussed above. Such 2-dimensional representations are usually downscaled to avoid computational overheads. Nataraj et al. recommended the dimensions to downscale 2-dimensional representations according to the size of the original file in [65]. Another approach by Chen et al. in [68] provides a similar recommendation but with a linearly scaled relationship between pixel file size (a multiplier of file size) and the image width. Both recommendations are provided in Tables 5 and 6.

**TABLE 5.** File size vs 2D image width - recommended by [65].

| File Size (KB) | Image Width |
|---|---|
| 0 to 10 | 32 |
| 10 to 30 | 64 |
| 30 to 60 | 128 |
| 60 to 100 | 256 |
| 100 to 200 | 384 |
| 200 to 500 | 512 |
| 500 to 1000 | 768 |
| >1000 | 1024 |

**TABLE 6.** Pixel file size vs 2D image width - recommended by [68].

| Pixel File Size | Image Width |
|---|---|
| 0 to 10 | 32 |
| 10 to 30 | 64 |
| 30 to 60 | 128 |
| 60 to 100 | 256 |
| 100 to 200 | 384 |
| 200 to 1000 | 512 |
| 1000 to 1500 | 1024 |
| > 1500 | 2048 |

*b: COLOR IMAGE*

It is an extension of the grayscale representation described above, where conversion to a colored format is done by extending grayscale values to RGB channel values using tools like BinVis,[1] as done by Nguyen et al. in [40].

### 3) 3-D SEQUENCE
*a: LATENT PROJECTION*

Unlike dynamic analysis, the use of three-dimensional projection of latent information is still largely unexplored for static analysis. Abdelsalam et al. [69] employed a 3D-CNN model to tackle a mislabeling problem in a virtual machine (VM) environment, where performance metrics collected over a specified time interval are treated as model input.

### D. INTERDEPENDENCE

The features for static analysis discussed so far dealt with the structural properties of an ELF binary, its code-level properties, and its section and segment-level components. All of them are obtained from within the binary, hence, treated as 'Intra-file' properties. The 'interdependence' deals with the properties that are external to the binary and is concerned about its proximities with the surrounding environment.

**File Reputation:** The problem of identifying the reputation score of a file based on different external metrics to validate its maliciousness is discussed here:

*b: FILE TO MACHINE RELATION*

It represents the absolute or relative path information of a binary file which could provide contextual information with the capacity to reveal benign or malicious intents [70]. For instance, Dropper malware or Adware would find themselves in a directory that potentially increases interaction with the end-user [70].

*c: FILE TO FILE RELATION*

It deals with the influences that a file inherits directly or indirectly from co-occurring files in the environment [60], [71], [72]. The variations in the importance of such relations to a malware file as opposed to a benign file help isolate malicious files. Ye et al. [60] explored the combination of file-to-file relation and file content in their work.

### E. DYNAMIC ANALYSIS

Dynamic analysis is not in the scope of this survey paper. However, we provide a high-level taxonomy of dynamic features observed in the literature that can be categorized into traced-based and usage-based features.

### 1) TRACE-BASED
These features deal with acquiring knowledge about malware activities and interactions over a period of time, such as tracing the API calls made by malware, tracing the sequence of instructions they executed, and their network interactions.

### 2) USAGE-BASED
These features deal with monitoring the usage of system resources such as memory, registers, and file access.

### 3) FILE PLACEMENT
It is a specialized file monitoring technique where decoy files are placed in suspected locations of malware activity. For instance, when ransomware tries to access the decoy file to steal information, its file system activity and access behavior are recorded for taking remedial actions in reality [73].

## IV. DISCUSSION ON SURVEYED PAPERS

In this section, we summarize the research works surveyed, and the significant findings reported in those studies, the advantages and limitations of their approach. The methodology used for selecting research works for our survey includes: i) consideration of the research works on IoT malware in the past five years, ii) inclusion of the prominent research works on Linux malware in the past ten years. Figure 7 provides a timeline of the surveyed works, and Table 7 provides an overview of the datasets used by them. A high-level comparison of surveyed works is provided in Table 8 following the discussion.

**Alasmary et al. [41]** proposed a control flow graph-based detection of IoT malware. They compared the complexity of IoT malware with Android malware and found that the former possessed higher complexity and rich structural flow. Their study used 2,999 benign samples and 2,962 malware samples for IoT collected from GitHub source codes and CyberIOC, respectively. They analyzed different graph properties, including closeness, betweenness, degree, shortest path, density, count of nodes, count of edges, and extracted 23 features to perform detection. Their study reveals that the CFG with the analyzed properties can significantly reveal evasion techniques like packing and obfuscation.

**Alhanahnah et al. [21]** proposed a cross-arch work inspired by software defect detection approaches. They used structural, statistical, and string features of high-level code to develop lightweight cross-architecture invariant IoT malware signatures. Their study used 5,150 samples from the IoTPOT dataset containing samples compiled for eight different CPU architectures and employed a multi-stage clustering followed by a YARA-based signature generation scheme to label IoT threats. The multi-stage clustering comprises initial K-means coarse-grained clustering using eight statistical features from code, fine-grained clustering using high-level graph structural similarity aided by a tool called 'BinDiff,' and then a final cluster merging phase based on N-gram similarity. High-level

---

[1]github.com/cortesi/scurve/blob/a59e8335c48a7cda7043fbd1b28bcae1abc9645d/binvis

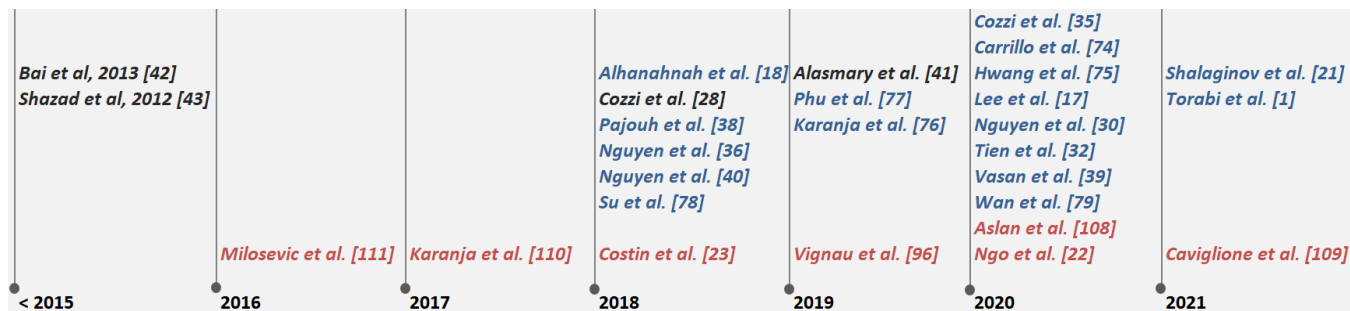

**FIGURE 7.** Timeline of research works on cross-architectural IoT malware threat hunting (highlighted in blue). Works highlighted in red represent general/cross-architectural IoT related survey works. Other important works are highlighted in black.

string statistic features are generated from merged clusters to form efficient signatures.

The study claims that the high-level code statistic features used by the initial coarse-grained clustering phase are resilient to cross-architecture variations, as they abstract away the different code syntax. The features used are: Functions count, Instruction count, Redirect instructions count, Arithmetic instructions count, Logical instructions count, Transfer instructions count, Segments count, count of call instructions. Their fine-grained clustering phase evaluates pairwise code structural similarity among samples within a coarse cluster, using techniques like single-linkage hierarchical clustering to generate dendrograms, and uses the Bindiff tool to reconstruct CFG abstracting structural features - ignoring assembly level features. The limitation of this work is that their study is based on a minimal number of samples compared to the magnitude of real-world samples and the proposed approach relies on NxN similarity score as a feature set that grows quadratically with dataset size - leading to insufficient time and resource efficiency.

The study also indicated the existence of cross-architectural similarity if the samples are from the same malware family. Like Hwang et al. [75], their study also asserts using Strings as a better cross-architecture feature as they are easy to extract. In addition, due to the intent of quick time-to-market, malware authors more often compile the same malware code to run on different CPU architectures, leading to easy identification of malwares as the string information stays the same even when underlying opcodes and other CPU specific instructions are different. Obfuscation and encryption techniques may randomize a human-readable string, yet the result is still a string that gets propagated across multiple CPU-specific malware clones, allowing them to be easily identified as their occurrences overlap across samples. Such overlapping information could act as robust features for cases like malware family classification. For instance, they found many of a malware family's samples to contain the same obfuscated sequences "eGAIM aJPMOG qCDCPK oMXKNNC uKLFMUQ."

**Bai et al. [42]** introduced a method that extracts system call information from the symbol table of ELF files and applied four machine learning algorithms for Linux mal-ware detection. Using a dataset from VXHeavens consisting of 756 benign and 763 malware samples (from 8 classes), the experimental results achieved more than 98% accuracy. They created a custom ELF parser as many parsers, such as readelf, failed to extract many samples. Frequency of system calls are used as the feature set characteristics, and they used the top 100 system calls obtained via ranking using information gain for training and classification. The authors claim that their system calls-based method is hard to evade even with obfuscation, polymorphism, and metamorphism.

**Carillo Mondéjar et al. [74]** proposed a data-driven approach for automating the study of IoT malware and extracting features for characterizing malware relationships that could aid in the IoT malware attribution/vetting process. They used such identified static/dynamic for features to identify unknown malware in the wild via similarity-based clustering. Interestingly, there were nearly zero singletons found in the unknown samples they analyzed for different processor architectures such as ARM, MIPS, and PowerPC, and they were able to associate the samples to at least one known family. They showcased how some malware families are more prevalent in certain processor architectures. For example, ARM suffers from more 'Mirai' botnets, while PowerPC suffers more 'Gafgyt.'

**Cozzi et al. [28]** proposed the first detailed Linux-specific malware analysis pipeline involving analysis of file metadata, static and dynamic features. For their comprehensive Linux malware study, they used a dataset containing a total of 10,548 samples from 10 different architectures, with sizes ranging from 134 bytes Backdoor to 14.8MB botnets.

To improve the understanding of Linux malware, they analyze and provide insights into various behaviors exhibited by Linux malware. For instance, they found that DDoS attack BotNets dominate the Linux malware landscape, spreading over 25 families. This was attributed to the fact that their source code is available publicly and attackers often harvest poorly protected IoT devices to join large remotely controlled botnets. The study also reveals that ELF samples generally lack OS compatibility information and that fields such as OS/ABI (application binary interface) in the ELF header are rarely used, obscuring the knowledge on the target environment. The use of arbitrary linking and loading libraries

**TABLE 7.** Comparison of datasets employed in surveyed works.

| | Benign # | Malware # | Benign Source | Malware Source | CPU Arch. Covered |
|---|---|---|---|---|---|
| Alasmary et al. [41] | 2,999 | 2,962 | GitHub | CyberIOC | - |
| Alhanahnah et al. [21] | 130 | 4,000 (out of 5,150) | openwrt | IoTPOT | ARM, MIPS, Intel 80836, PowerPC, x86-64, Renesas SH, Motorola 68020, SPARC |
| Bai et al. [42] | 756 | 763 | Linux OS | VxHeavens | - |
| Carrillo et al. [74] | - | 10,548 | - | Cozzi et al. [28] | ARM 32, MIPS I, Intel 80386, PowerPC, AMD x86-64 |
| Cozzi et al. [28] | - | 10,548 | - | VirusTotal, Self | ARM 32-bit, MIPS I, Intel 80386, x86-64, PPC, Motorola 68k, SPARC, Hitachi SH, AArch64 (ARM 64-bit) |
| Cozzi et al. [35] (IoT Genealogy) | - | 93,000 | - | VirusTotal, Self | ARM 32-bit, MIPS I, PPC 32/64-bit, SPARC, Hitachi SH, Motorola 68k, Tilera, ARC, Interim Value tba |
| Hwang et al. [75] | 10,000 | 10,000 | Linux OS | VirusShare | - |
| Karanja et al. [76] | 143 | 258 | Linux OS | IoTPOT | MIPS, x86, SUPERH |
| Lee et al. [20] | - | 122,504 | - | VirusTotal | ARM, MIPS, X86, X86-64, PPC, SPARC |
| Ngo et al. [22] | 4,001 | 7,199 | IoT SOHO Firmwares (binwalk tool) | IoTPOT, VirusShare | No information |
| Nguyen et al. [40] | 1,000 | 1,000 | Ubuntu 16.04 | IoTPOT | x86 (ignored ARM, MIPS) |
| Nguyen et al. [36] | 6,031 | 4,002 | SOHO IoT devices | IoTPOT | ARM, MIPS, PPC, SPARC |
| Nguyen et al. [30] | 3,845 | 6,165 | Ubuntu 16.04 | IoTPOT, VirusShare | ARM, Intel 80386, PPC, MIPS, SPARC, Motorola m68k, X86-64, SuperH |
| Pajouh et al. [38] | 271 | 280 | Rasberry Pi II compatible Linux Debian packages (pkgs.org) | VirusTotal | ARM (32-bit) |
| Phu et al. [77] | 4,107 | 19,710 | Ubuntu OS - x86-64, Intel 80386. Firmware router images Zyxel, TP Link, Belkin, Openwrt, Asus, Dlink etc | IoTPOT, Detux, VirusShare | MIPS, PowerPC, SPARC, ARM, Motorola, RenesasSH and PC |
| Shalaginov et al. [8] | - | 10,574 | - | VirusShare | Intel binaries |
| Shazad et al. [43] | 734 | 709 | Linux OS (no version specified) | VXHeavens, Offensive Computing | No information |
| Su et al. [78] | 365 | 365 | Ubuntu 16.04.3 | IotPOT | No information |
| Tien et al. [32] | 2,157 | 6,251 | Ubuntu 16.04 LTS | CZ.NIC, IoTPOT | x86, x64, ARM |
| Torabi et al. [1] | - | 74,429 | - | VirusTotal, VirusShare, IoTPOT | - |
| Vasan et al. [39] | 5,655 | 15,482 | Firmware router images - Asus, Belkin, Tenvis, Dlink, TP Link, Linksys, Trendnet, Centurylink, Zyxel and Openwrt | IoTPOT, Detux, VirusShare | ARM, MIPS, Intel 80386, x86-64, PowerPC |
| Wan et al. [79] | 111,353 | 111,610 | D-Link, Zyxel, Netgear, IDIS, Belkin, MikroTik | VirusTotal | ARM, MIPS, x86, x86/64, PowerPC, SPARC, Renesas SH |

like glibc, uclibc, libpcap, libopencl, musl, also complicates malware analysis.

**Hwang et al. [75]** attempted to avoid the usage of structural information from executables. Instead, they used string information from platform-independent binary data. They aimed to create a single auto-analysis mechanism for malwares from different OS platforms such as Windows/Linux/IoT malwares. Their analysis method involved getting raw binary samples via device endpoints, extracting around 1800 different strings including API names, DLL names, library function names of programming languages,

PE/ELF file formats, converting them to hashed format, employing count based string vectorization on the obtained hashes, performing ExtraTreeClassifier based feature selection to find the most useful string hashes, and finally employing KNN based detection.

Unlike [32], they closely indicate that the most used Linux OS for IoT devices is Ubuntu Core. The major limitation of their work is that their method may not be effective over packed or encrypted samples because their method depends only on the strings available explicitly, and such packed/encrypted samples reveal limited string information during static analysis. The authors claim that string-based techniques are better candidates for cross-architecture malware detection/classification as they are more resistant to obfuscation. Interestingly, they were successful in parsing more malwares than benign samples, but they failed to indicate the type of parser used and the reason for failure in benign sample parsing other than indicating that there are different file formats within ELF.

**Karanja** *et al.* **[76]** proposed a classification approach based on textural features extracted from IoT samples. The samples were first converted to a grayscale representation. Then a gray-level co-occurrence matrix, shortly 'GLCM,' was generated, which are used to calculate five of the Haralick features proposed by Haralick *et al.* [80], namely: Correlation, Inverse Differential Moment (IDM), Entropy, Angular Second Moment (ASM), Contrast. The authors claimed that such textural features possess the capability of processor architecture and OS platform independence for the classification task and are low in computational requirement due to the use of such image representation that could be generated and analyzed for samples from any platform. They failed to provide clear evidence by accounting for the evaluations and results over different processor architectures.

**Lee** *et al.* **[20]** made use of 122K samples from VirusTotal covering various architectures like ARM, MIPS, x86, x86-64, PowerPC, SPARC, and extracted printable string information such as the plain texts from function names, API names, codes, and code comments. The distinctive contribution of this work is that they studied the effect of learning samples from broadly used architectures and then evaluate the model performance on lesser-known architectures as an unseen testing set, and their models performed better in such cases.

**Nguyen** *et al.* **[40]** proposed "BE-PUM" - Binary code analyzer based on dynamic symbolic execution on x86 – that overcomes obfuscation for precise malware disassembly. IoT malwares do not employ obfuscation as frequently as PC malware. Their dataset contained 1000 IoT malwares for x86 provided by the IoTPOT team, Yokohoma national university. The study compared three different models built based on CNN but employing different types of features.

The feature types compared include assembly instruction sequences extracted using ObjDump tool as features and trained with CNNs, fixed-size byte sequences learned using Malconv CNN model, and fixed-size color images learned using AlexNet CNN obtained using Hilbert Curver images from binary code entropy. The color images were generated using Shannon Entropy on the entire byte sequence and synthetically extracted information for the RGB channel. The study discusses findings such as the average size of IoT malwares being around 1MB, while the benign samples' average size is around 0.07MB. LSTM models work well for samples less than 0.5MB. A limitation of this study is that they claim to allow variable-sized instruction sequences for processing by CNN models, but no evidences are provided.

**Nguyen** *et al.* **[36]** proposed an approach combining printable string information (PSI) graph and deep graph convo-lutional neural networks (DGCNN) for the task of detecting Linux-based malwares. Their dataset comprised 6K benign samples from small office/home office (SOHO) and 4K malware samples collected from IoTPOT [81]. Their study deals with learning IoT malwares from different architectures such as ARM, MIPS, PowerPC, and SPARC, leveraging the linkage among printable strings.

**Nguyen** *et al.* **[30]** proposed a lightweight printable string information-based graph (PSI-Graph) for the detection of botnets in IoT, as an extension of their work above [36]. The PSI-Graph is meant for capturing the relationship structure among the functions consisting of PSI. The PSI-Graph extraction is done as a simplified version of the function call graph (FCG) using only the functions that contain PSI characters with lengths more than or equal to three. The dataset they employed consisted of a total of 11.2K ELF files, among which there were 6,165 malicious botnets and 3,845 benign samples. Though their method improves on the drawbacks of FCG, they still suffer from a very high prediction time per sample at the rate of 1,140 ms/sample (88 minutes for 4,630 samples).

**HaddadPajouh** *et al.* **[38]** proposed a deep LSTM based approach for dealing with IoT malwares. Their dataset consisted of 32-bit ARM-based malwares with 280 malwares and 271 benign samples from VirusTotal. They decompiled ELF files using the ObjectDump tool, extracted more than 600 Opcodes, performed binary encoding, and then pruned them based on TF-IDF frequency and the number of occurrences, followed by further feature reduction using PCA. Operands and registers info are also pruned out, and only the opcodes are considered.

**Phu** *et al.* **[77]** used the Valgrind tool's intermediate representation called 'Vex' and 'CFD' - a dynamic programming-based extraction of control flow-based features that combine opcodes and control flow graphs to build a cross-architecture malware detection system. While Vex provides architecture-agnostic representation support, CFD helps in achieving better malware detection performance.

Interestingly, they found that training the CFDVex model on Intel 80386 samples performed better on the testing set with only MIPS samples. However, vice versa was found to be performing poorly.

**Shalginov and Øverlier** *et al.* **[8]** focused only on Intel-compatible Linux malware (desktops, servers), and IoT environment-specific malware like ARM binaries are not considered. The study claims to overcome the limitation of static features in multinomial Linux classification due to obfuscation, but no clear pieces of evidence were provided. The study used a dataset with a total of 15,101 Intel-compiled ELF samples from VirusShare (one Linux archive and three ELF archives). They filtered 10,574 malware files belonging to 442 malware families such as Tsunami, Mirai and 19 malware types such as Trojan and Worm. The features they extracted include Virustotal report, preframe report, readelf, strings, file size, and file entropy.

The study indicates that the Intel platform has 1128 file types differing in 32/64-bit binaries, LSB, static/dynamic linked, stripped, and processor version. They also indicate that Linux machines are often infected for distributed attacks and botnets like Mirai, DDos. Also, the study recommends using CARO (Computer Antivirus Research Organization) for malware family labeling, which cybersecurity companies like TrendMicro are following.

**Shahzad and Farooq** [43] extracted structural features from ELF binaries to train malware detection classifiers. The authors attempted to explore another dimension of malware research that whether using such structural information to detect malware, as shown for Windows PE files in [82], could be generalized to executables of other OS platforms such as Linux. Their work does not account for different CPU architectures, and the gathered features are more prone to evasion techniques such as packing and binary stripping.

They used a dataset collected from VX Heavens and Offensive Computing, with 734 benign and 197 malware samples having sizes ranging from 20KB to 4MB. They extracted a total of 383 structural features, and interestingly much of the features were from Section header than ELF header and Program header combined. They used measures such as Resistor-Average (RA) Divergence – based on KL-Divergence Frequency Histogram (Over Sections) to evaluate their finding that the structural information of malware samples differ significantly compared to benign samples.

Some of the significantly different ELF structural features include ELFHEENTRY, ELFHEPHNUM, ELFHESH-NUM, and ELFHESHSTRINDX fields in the ELF header, and there are frequently occurring sections such as.comment,.note,.strtab,.symtab, and.sbss in malware, and.rel.dyn, got.plt in benign samples. They also discussed the significance of PC-Relative addressing for relocation: RELR386PC32. Malware tends to use relative addressing, while absolute addressing is common in both benign and malwares. They also discussed that most malware misuse.comment,.notes sections (used to hold version control information and file notes, respectively) to store their malicious information.

**Su *et al.*** [78] attempted to perform a lightweight classification of IoT DDoS malware. Their dataset comprised only around 365 samples for benign and malware each. The limitation of this study is that they downscaled malware binaries to a $64 \times 64$ grayscale image representation irrespective of the size of the actual malware. Hence, as the dataset population grows, the downscaled images may not hold adequate information for their model to learn, and their method's performance may degrade faster on unseen samples.

**Tien *et al.*** [32] discussed how IoT malware detection approaches are different from Windows PE malware detection. Conventional anti-virus techniques based on the Windows paradigm cannot be applied directly to counter threats to the IoT devices due to the tight dependency that architecture-specific feature types induce in the training dataset. Such issues call for a universal feature representation via static analysis to be able to achieve OS platform independence and cross CPU architecture applicability.

They proposed two feature groups for such purpose in their study: feature group 1 - dealing with seven types of static meta-features of ELF samples - ISA (8 categories), sample size, external/shared library (ldd command), packer, number of functions, potential behaviors via YARA (anti-VM detection, exploit-kit attacks), networking (gauged by IP address strings and protocol function names – not by execution); feature group 2 - 12 types of Opcode instruction features obtained via reverse engineering. They collected 30,146 malwares - 25,000 samples from the IoT HoneyPot project Yokohama National University and 5,146 samples from CZ.NIC.

The major disadvantage of this cross-architectural study is that they have included only the samples for x86, x64, and ARM, totaling 6,251 out of 30,146 samples, but ignored samples from other architecture types. Including such samples would introduce data imbalance (due to non-uniformity in public sample availability) and makes the malware detection problem more challenging under this scenario - which is a valuable research direction towards ISA neutrality. Metadata for the 6,251 samples have also been released by the authors.[2] Another disadvantage is that their dataset is biased. Their benign samples were collected from Ubuntu 16.04 LTS Client for PC-like devices, but malware samples are from IoT devices. Though the authors claim that both have enough common features, they failed to provide more information on how similar/different they are in practice.

Since their dataset consisted of samples compiled for different CPU architectures (x86, x64, ARM), the opcodes extracted from those samples would have different instruction names but may perform the same operation. Hence, they grouped the extracted opcodes into 12 groups based on the functionality, namely: Logic, Control, and Status, Memory, Stack, Procedure, Prefixed, SystemI/O, Arithmetic, System, Branch, Execution Time, and Others. Among these groups, the Memory, Arithmetic, Branch, and Other opcode groups are found to be more beneficial for both cross-architectural malware detection and family classification than the rest. Another finding that closely resembles IoT malware characteristics is that most malware samples exhibited networking invocation-related strings and IP addresses than benign samples in their dataset. However, it may be a biased finding since their benign samples are collected from PC-based OS.

**Torabi *et al.*** [1] proposed a malware classification system based on multi-modal learning via deep learning approaches. Their method followed a feature fusion approach with LSTM and CNN models used over the string and grayscale information extracted from the samples, respectively. Their dataset comprised 70K IoT malwares, and they showcased that unknown malware labeling could be done using DL approaches and that IoT malware obfuscation is not sophisticated. They were able to de-obfuscate 76% of the strings from

---

[2]drive.google.com/open?id=1NXNsPYa2Qz0cZEy3u3bn_Kx4SVUrykAz

**TABLE 8.** Comparison of features employed in surveyed works.

| Work By | Is Cross-Architecture? | Task | Feature Representation | ML Approach | Feature Selection /Reduction | Benefits | Weakness |
|---------|------------------------|------|------------------------|-------------|------------------------------|----------|----------|
| Alasmary et al. [41] | No | Detection | Control flow graph (CFG) | SVM, CNN, LR, RF | None - as only 23 features were used | Graphical properties of CFG could reveal existence of packing and obfuscation | Different IoT architectures are not taken into account. CFG extraction is generally time consuming and prediction time complexity is not discussed. |
| Alhanahnah et al. [21] | Yes | Detection Classification | String based N-gram, structural, and statistic features, YARA rules | K-means clustering, Similarity analysis | Bindiff based graph structural pair-wise similarity, String similarity, Jaccard similarity (for unseen samples) | Cross-architecture resilience as code statistic features abstract away code level syntax. Lightweight signatures. Resource and energy efficiency. Resilient to obfuscated strings. | Only around 5k samples are used. Lightweightness may be impacted as population grows. Energy efficiency properties are not benchmarked. |
| Carrillo et al. [74] | Yes | Classification | n-gram of opcode sequences | k-NN, DT, RF, SVM | Cyclomatic complexity based static feature selection and behavioral dynamic features including User and Group Identifier, number of processes created, process renames, I/O controls and unique syscalls | Malware relationship characterization especially for unknown malware | Samples not labeled by AVClass tool are ignored. This majority of AMD x86-64 samples in the considered dataset are left out. |
| Cozzi et al. [35] | Yes | Classification | HNSW - Hierarchical Navigable Small World lineage graphs for Phylogenetic trees (malware evolution) | Function level similarity graph clustering | A Cut measure to eliminate libraries functions belonging to standard libraries and isolate user defined functions | Helps in identifying mislabeled samples. Provides knowledge on Intra family evolution and relationships by code reuse. | Intel and AMD samples are ignored. Proposed symbol propagation method is limited to symbols identified in unstripped binaries, i.e., not all symbols in stripped binaries are matched. |
| Hwang et al. [75] | Yes | Detection | Uses count based string hash vectorization to represent extracted string info like API, DLL, library names | k-Nearest Neighbor | ExtraTreeClassifier based feature importance | Avoids structural information from binary as it is platform dependent. | Depends only on the strings available explicitly - limited application over packed samples |
| Karanja et al. [76] | Yes | Classification | Grayscale Images, Textural features extracted using Gray Level Co-occurence Matrix (GLCM) by Haralick et al. [80] | k-NN, NB, RF | 5 out of 14 Haralick features: Correlation, Inverse Differential Moment (IDM), Entropy, Angular Second Moment (ASM), Contrast | Platform independent feature via selected Haralick textural features | Cross-architecture efficiency is not clearly discussed |
| Lee et al. [20] | Yes | Classification | String length frequency (SLF), Printable string information (PSI) from code comments, function & API names | RF, k-NN, SVM | Recursive Feature Elimination (RFE) via SVM, Document Frequency (DF) based elimination | First study to learn on major IoT architectures and test on lesser known architecures | Assumed maximum SLF feature length is biased as it was based on complete dataset instead of training data |
| Ngo et al. [22] | Yes | Detection | ELF header, Strings, Opcodes, Images, Control flow graphs (CFG), Printable string information (PSI) | RIPPER, PART, DT (J48), SVM, k-NN, RF, NN, LR | No information | Discusses interaction between known botnet families (IoT) | Shallow taxonomy of feature representations. No details about architectures of samples in the benchmark dataset used for comparison of studies. |

**TABLE 8.** *(Continued.)* Comparison of features employed in surveyed works.

| Work By | Is Cross-Architecture? | Task | Feature Representation | ML Approach | Feature Selection /Reduction | Benefits | Weakness |
|---|---|---|---|---|---|---|---|
| Nguyen et al. [40] | Only x86 | Detection | Byte sequence, Assembly instruction sequence and Color image representations | CNN | Abstraction of opcode parameters i.e., operands and registers to reduce dimensionality | Uses - BE-PUM, a binary code analyzer for precise malware disassembly by overcoming obfuscation | Claims to allow variable-sized instruction sequences for processing by CNN models but no information provided |
| Nguyen et al. [36] | Yes | Detection | PSI-Graph | Deep Graph CNN (DGCNN) | Functions with length of PSI characters length at least 3 are only considered | Accounts for PSI contexts by leveraging the linkages among printable strings | Prediction time model efficiency is not discussed |
| Nguyen et al. [30] | Yes | Detection | PSI-Graph from functions with printable strings via function call graph (FCG) | CNN | Functions with PSI characters of length at least 3 are only considered | Minimizes the complexity of using function call graphs by reducing to PSI-graph | PSI-graph still suffers from high prediction time per sample (more than 1,000ms) limiting its practical use |
| Pajouh et al. [38] | Only ARM | Detection | TF-IDF frequency, Number of occurrences and binary encoding for Opcode sequences | RNN, LSTM, BNN (Bi-directional neural network) SVM, RF, NB, MLP, k-NN, AdaBoost, DT | Information Gain (IG), Principal Component Analysis (PCA) | First study to adapt deep learning to IoT malware detection | Very small dataset. Only ARM based binaries are used. Only opcodes are considered, operands and registers info are pruned out |
| Phu et al. [77] | Intel 80386, MIPS | Detection | CFD - Control flow based features with Vex as intermediate representation | SVM | Chi-square | Use of heterogeneous training and testing sets for evaluating cross-architecture detection efficiency | Claims to have used the largest IoT dataset, yet only Intel 80386 and MIPS architecture samples are used in the study |
| Shalaginov et al. [8] | Only Intel ELF binaries | Classification | Structural and entropy information from readelf, strings, file size, file entropy, strings, VirusTotal report, peframe report | Deep Neural Network, Weka 3.8.4 for NB, SVM, MP, k-NN, C4.5 | Information Gain (IG) | Overcomes limitation of static features in multinomial Linux classification due to Obfuscation | Focus is only on Intel compatible Linux Malware (Desktops and Servers) - ARM, MIPS and other architecture binaries are not considered |
| Shazad et al. [43] | Yes | Classification | ELF structural information: ELF header, Section header, Program header, symbols, dynamic symbols, dynamic section, relocation sections, global offset table, hash table | Rule-based classifiers - RIPPER, PART, C4.5, DT, J48, and Bio-inspired classifiers - cAnt Miner, UCS, XCS, Gassist | Information Gain (IG) | Adapted structural information-based malware detection approach from [82] | Does not account for different architectures. Prone to easy evasion – binary stripping / packing |
| Su et al. [78] | No | Detection Classification | Gray-scale images | CNN | Down Scaling (64 x 64 image) | Lightweight classification scheme | Proposed scheme is no longer lightweight when large samples have to be dealt |

**TABLE 8.** *(Continued.)* Comparison of features employed in surveyed works.

| Work By | Is Cross-Architecture? | Task | Feature Representation | ML Approach | Feature Selection /Reduction | Benefits | Weakness |
|---|---|---|---|---|---|---|---|
| Tien et al. [32] | Yes | Detection Classification | Opcodes and ELF static features: Instruction set architecture (ISA), sample size, external/shared library, packer, number of functions, potential behaviors via YARA, networking metrics | ANN, CNN, SVM | Opcodes of different ISAs are grouped into 12 types to achieve ISA neutrality and reduce features dimension: logic, control and status, memory, stack, procedure, prefixed, system input and output, arithmetic, system, branch, execution timing, and the other least occurring opcodes | Adapted opcodes to improve cross-architecture detection and classification | Opcode extraction using IDA Pro may be limited by anti-disassembly techniques. Limitation in analysing payload of "dropper" malwares for Opcodes. Limited ISA neutrality as filtered dataset contains x86, x64, ARM only. |
| Torabi et al. [1] | No | Classification | Grayscale images, Opcodes | LSTM, CNN | Top 50K strings | Evaluated their DL approach to label 24K unknown samples. Analysis on evolution of Mirai during Covid-19 pandemic | Claims real-time performance but achieved prediction time of fusion model not discussed |
| Vasan et al. [39] | Yes | Detection | Opcode sequence | RNN, CNN | Opcode dictionary and Information Gain (IG) | Proposed opcode dictionary approach helps to create a compact representation and minimize the data dimensionality to a greater extent. Detects obfuscated malware – including metamorphic and polymorphic | Prediction time of proposed model is 0.32s (320ms) which is still high. The representative dataset used for evaluation is small, and the proposed approach suffers from a false positive rate which is very high. Use of Information Gain on opcode sequence to train CNN is not explained clearly |
| Wan et al. [79] | Yes | Detection Classification | N-gram from byte sequence at ELF's entry points | SVM, NB, KNN, MLP | sparse n-gram representation. No other information | Promotes cross-architecture performance by considering CPU-specific information for learning | Entry points based byte sequence may not provide sufficient discriminating data in the presence of code reuse among malware families |

samples they examined using FLOSS [83] as the samples used only standard UPX packers.

**Vasan et al. [39]** proposed a stacked ensemble learning employing both RNNs and CNNs, with heterogeneous feature selection algorithms based on opcode dictionary and information gain (IG). It was able to detect obfuscated malware as well as metamorphic and polymorphic malware. Their dataset consisted of 21,137 samples, including 5.5K benign and 15K malware samples. They showed that their approach is robust against junk code and benign opcode insertions.

**Wan et al. [79]** proposed a detection and classification approach utilizing the n-gram byte sequences near the entry points of the binary samples. They made use of a dataset with 111K samples each for both benign and malware categories and showed that there are merits in considering CPU-specific information to achieve better cross-architecture performance. However, their method performs poorly in the presence of code reuse among malware families.

## V. CHALLENGES AND RESEARCH OPPORTUNITIES
### A. CHALLENGES
Generally, for malware threat hunting, there is no 'one model fits all' type of solution that learns all feature representations and detects all threats. In the case of Windows threats, even though there exist several mature approaches, millions of threats are still left undetected. Adding fuel to the fire, there are zero-day attacks, and the problem of concept drift - machine learning models also suffer from continuous decay of learned matter. On top of these issues, the IoT malware threat hunting problem spans multiple folds across other issues like the need to tackle a wide variety of malware families, a wide range of OS platforms, and numerous CPU architectures. Table 9 provides an overview of the challenges discussed in this section.

### 1) DATASET AVAILABILITY

There exists a severe lack of representative datasets for ELF when compared to Windows PE malware datasets at present since Linux malware gained the attention of cybersecurity only around 2014, much later than PE malwares. For instance, there are many public benchmark datasets like the EMBER PE feature-based dataset [84], and the recent industrial-scale dataset called SOREL [85], and the temporal PE dataset BODMAS [86]. Recently, IoTPOT team released an extended dataset comprising 173K IoT malware binaries [87], which is the only known large benchmark IoT dataset to date.

Despite massive efforts from providers like VirusTotal, VirusShare, there exists a severe imbalance in the number of samples available under each type of CPU architecture and OS platforms. In such complex environments with various devices, vendors, architectures, and commands, it is challenging to construct large-scale datasets [23], [75], leading to the class imbalance between different malware families and the class imbalance between malware samples compiled to different ISAs.

### 2) BENIGN BINARIES AVAILABILITY

On top of the lack of datasets discussed above, it should be noted that collecting benchmark benign samples are much harder. Datasets built on vectorized feature representations include the data for benign samples, however, datasets sharing raw binaries do not include benign samples, as a consequence of adherence to software license agreements that prevents the authors from sharing the actual samples. Such incomplete datasets limit further feature explorations.

### 3) ENVIRONMENT FOR ANALYSIS

Linux cross-architectural malware analysis environment is challenging to set up with a compatible set of OS/distros, processor architectures, and supporting libraries [28], [75].

### 4) EVASION TECHNIQUES

Static analysis challenges such as Code obfuscation, Code Encryption, Payload dropping are also applicable for IoT malware. However, IoT malware in the wild is not yet as sophisticated as Windows malware as they also need to tackle various constraints discussed above [1], [21], [30], [79].

### 5) HARDWARE CONSTRAINTS

Most of the IoT devices are never turned-off during their lifetime for continued operation, and they are equipped with limited resources such as memory and storage capacity. Unlike PCs with AMD/x86 architecture, their computing power is very minimal due to such constrained hardware resources [90].

**TABLE 9.** Overview of challenges.

| Challenge Description | Reference |
|---|---|
| Dataset availability | [81], [85], [86], [88], [89] |
| Benign binaries availability | - |
| Environment for analysis | [28], [75] |
| Evasion techniques | [21], [30] |
| Hardware constraints | [90] |
| Lack of device standards | [75], [91], [23], [28] |
| Lack of security standards | [92] |
| Lack of softare standards | [23], [75] |
| Lack of studies | [28], [75] |
| Linking and Loading | [75], [28] |
| Malware family diversity | [8] |
| Model complexity | - |
| Multiplicative effect of malware | - |
| Open-source implications | [93], [94] |
| Open-source Tools | - |
| OS diversity | [28], [75] |
| Packing issues | [95] |
| Processor architecture heterogeneity | [28], [75] |
| Real-time evaluation | [23], [75] |
| Role of emulators | [90] |

### 6) LACK OF DEVICE STANDARDS

The vulnerability of the IoT devices grows exponentially as they become more and more interconnected, and the risk of getting hacked grows with it [91]. Managing IoT devices and their data is already a challenging big data problem. If one of the systems in the network gets compromised, other devices are at higher chances of risk of getting hacked/corrupted. There also exist challenges in the communication between devices manufactured by different IoT device vendors, as there currently exists no international IoT compatibility standards [91].

Lack of OS compatibility information makes it is hard to discern where the binaries are supposed to run. For instance, the 'OS/ABI' field in the ELF header is currently insufficient and rarely used. It is possible to specify different OS/ABI values than the compatible OSes for which they are compiled [28]. There is also difficulty in reliable extraction of data and metadata by identifying the correct firmware, which is a common challenge when dealing with a reverse engineering analysis of firmware [23], [75].

### 7) LACK OF SECURITY STANDARDS

Driven by the time-to-market necessities and the need for better usability and performance features, appropriate security and encryption mechanisms are being given low importance by the IoT device manufacturers. Due to these reasons, they suffer from being hacked massively, such as the D-Link routers case where several vulnerable routers were hacked and installed with backdoors that allow remote Command and Control for malware authors for use as cyber-attack weapons [92]. Also, most end-users of IoT device does not usually change the default credentials leaving them vulnerable to botnet attacks.

### 8) LACK OF SOFTWARE STANDARDS

Unpacking and decoding custom formats would be easy to address for standard software components, as they have standardized formats for the machine resources, code, and file organizations, but not for embedded software distribution, as it lacks standards [23], [75].

### 9) LACK OF STUDIES

Most of the studies for IoT malware threat hunting adapt existing Windows-based techniques and may not cover all types of processor architectures, as they typically focus either on general IoT malware or focus on prominently used architectures. It is also unclear how to collect balanced cross-architectural datasets, design and implement an extensive tailor-made analysis pipeline [28], [75].

### 10) LINKING AND LOADING

As highlighted by Hwang et al. [75], statically linked binaries may pose some challenges. Static linking is usually leveraged to achieve portability. However, analysis of such files is time-consuming. Further, the existence of arbitrary linking (glibc, uclibc, libpcap, libopencl, musl) and loading libraries necessitates the maintenance of a rich knowledge base up-to-date. In the case of dynamic analysis, if there is no appropriate loader and library in the analysis environment, malware samples would be prevented from starting execution and results in failure of analysis [28], [75].

### 11) MALWARE FAMILY DIVERSITY

A challenge is that there is no universal standard for malware family naming. Different vendors follow different naming conventions for storing identified malware samples, resulting in different names for the same sample. Cybersecurity companies like TrendMicro use CARO (Computer Antivirus Research Organization) naming scheme as their naming standard.

Another challenge is that there exist too many malware families, which could complicate the model learning process. For instance, the study by Shalginov & Øverlier [8] used the DNN model, and they discussed that it was hard to boost the classification accuracy of their DNN models by increasing the number of hidden layers when there are 10s to 1000s of malware classes (family classification scenario).

### 12) MODEL COMPLEXITY

In terms of cross-architecture IoT malware threat hunting, the data representations to learn models should be generated using lightweight cross-architecture signature generation schemes to improve the compactness and accuracy of models and to improve resource efficiency and energy efficiency for local deployment.

### 13) MULTIPLICATE EFFECT OF MALWARE

The infected zombie devices in botnets could allow remote management of affected devices and allow extortion of con-

*readelf:*

*radare2:*

**FIGURE 8.** Limitations of feature extraction tools.

fidential and sensitive data, or to recursively add the other connected devices as part of the botnet.

### 14) OPEN-SOURCE IMPLICATIONS

Open-source operating systems such as Linux, Android, and their variants cover most of the IoT market, and as their code is open-sourced, it makes the life of a hacker easier to analyze the available code and find vulnerabilities to exploit [90].

### 15) OPEN-SOURCE TOOLS

Tool level limitations also need to be considered. The rate of successful extraction of features from all the dataset samples being studied depends on the capability of tools employed. For instance, opcode extraction using IDA Pro may be limited by anti-disassembly techniques.

Figure 8 showcases an example where the readelf tool was able to successfully find the reason for not being able to parse a file successfully - due to corrupted ELF header, but radare2 failed to find the same.

### 16) OS DIVERSITY

Due to the existence of multiple custom OSes for IoT devices, there could be a number of issues with interoperability [28], [75], and software package compatibility.

### 17) PACKING ISSUES

Various works have been published to identify the family of the packer, such as the consistently executing graph-based approach by Li *et al.* [95]. Entropy also plays an essential role in identifying packing or encrypted binary blobs. Nevertheless, many custom packers need to be effectively studied.

### 18) PROCESSOR ARCHITECTURE HETEROGENEITY
There is much more number of different processor architectures in the IoT landscape compared to PCs and Laptops [28], [75]. Moreover, each has its own customized ISA.

### 19) REAL-TIME EVALUATION
A final activity in the machine learning process for IoT is to evaluate the model learned by deploying it in real-firmware devices. Such challenging task needs manual analysis and consumes more human effort, like the valuable bandwidth of malware analysts. It gets even more challenging to cover all the devices/images in the network [23], [75].

### 20) ROLE OF EMULATORS
Emulators are very much helpful in emulating different processor architecture environments, but the emulated systems may not be complete as running an actual firmware [93], [94].

Several other vulnerabilities are forming the base for the exponential increase in IoT attacks [21], such as the requirement of constant online connection by IoT devices to be smartly connected, lack of protection and security mechanisms, and their integration into human life makes them an exciting attack target.

### B. GAPS AND RESEARCH OPPORTUNITIES
Often suggested topics in existing surveys as research opportunities are explainability, adversarial learning, and advanced persistent threats (APTs) mitigation. We present a few of the gaps present in the existing literature that we identified and also other novel possible research directions towards exploring the unknown.

Most of the existing malware threat hunting approaches on ELF are adapted from the techniques belonging to areas such as natural language processing (NLP), image processing, and graph processing, that are initially proposed for learning Windows malware. Unique ELF-specific problems such as platform-independent malware threat hunting invited adaptation of approaches from other research areas like Software bug/defect detection [62].

The Linux malware study from Cozzi *et al.* [28] employed rolling entropy of Code and Data segments in general ELF files. Similarly, rolling entropy of sequences such as Assembly sequence, Byte sequence for IoT ELF could provide meaningful performance.

Lee *et al.* [20] proposed a cross-platform IoT malware family classification task based on printable string information (PSI). Their work suffered from the curse of dimensionality in terms of the vast number of unique PSI found. They lose valuable data by pruning PSI with frequency measures and recursive elimination. Approaches to reduce dimensionality without loss of valuable data could be explored.

A potential research direction is to leverage standard ELF features to support cross-architectural IoT malware threat hunting. For instance, Tien *et al.* [32] grouped similar opcode features based on functionality to make opcodes ISA independent. Other such valuable features could be identified and leveraged for better cross-arch performance. For instance, a grouping of processor-dependent structural information from the ELF header.

Further, the functionality-based grouping of opcodes by Tien *et al.* [32], such as arithmetic opcodes, branching opcodes, etc., serves a higher level of abstraction, meaning that twelve groups may not be enough to achieve superior performance on larger datasets. Instead, a granular grouping of opcodes based on their individual functionality would be more helpful. For instance, as in Table 3, opcodes with different names under different architectures for operations such as "Move" should be group together, instead of grouping them commonly under the high-level "Memory" functionality as done by Tien *et al.* However, such granular grouping of opcodes is highly-challenging, as many of the instruction set architecture specifications that are available for different architectures are obsolete or incomplete.

A gap in existing research works for ELF is that File-to-File relations, File-to-Machine relations, and their corresponding graph-based approaches are previously explored only for Windows PE binaries. Since the memory capacity and the software required to run the IoT device are very small when compared with Windows-based devices (like PCs, servers), it is interesting to study the effectiveness of such interdependence-based approaches in resource-constrained environments.

As a general guideline to affirm and better demonstrate the malware research works, we highlight the below as gaps in the literature structure:

#### a: EXTRACTION RATE
The extraction rate describes the success rate of the feature extraction technique associated with the proposed malware detection/classification approach. As described in Section V-A15, many of the feature extraction tools fail to achieve a 100% extraction rate, and the failed samples are ignored in general, which in turn results in incomplete learning and reduces the practical applicability of machine learning and deep learning approaches.

#### b: PREDICTION TIME
It is known that for practical usage of the model learned for malware threat hunting, they are expected to have a very short prediction time. This prediction time needs to be treated as an end-to-end time requirement; as such, it is the sum of the time taken for loading a sample from disk, performing feature extraction, pre-processing, and then the actual prediction. The end-to-end prediction time is expected to be on the scale of few milliseconds, say 10 ms, when carried out on a single-core CPU (not GPU). Therefore, the benchmarking information produced as part of experiments should include the end-to-end prediction time performance analysis.

### c: MEMORY REQUIREMENT

The machine learning/deep learning models are also expected to be lightweight, i.e., low memory footprint during execution, say 1 to 4GB. The memory needs of a model are influenced by factors such as the size of feature representations, the number of model parameters, and the libraries needed.

### d: STRINGENT FPR REQUIREMENT

False positive rate (FPR) is a critical measure for high stake domains like cybersecurity. It is essential to keep it well below 1% to reduce false alarms [67], which leads to the wastage of a significant amount of a malware analyst's valuable time for a proper investigation.

### e: BENIGN SAMPLES

As discussed in Section V-A2, to minimize the impact of lack of benchmark benign binary samples and improve reproducibility, the authors could release the SHA256 information of the benign samples used in their research work.

Finally, a helpful research direction is to contribute curated public benchmark datasets with binary samples and features for cross-architectural research, similar to public datasets being increasingly contributed for Windows PE recently: Ember [84], Sorel-20M [85], BODMAS [86], Virus-MNIST [88], MalNet [89].

### C. FORECASTS

The sheer scale of malware explosion along with millions of new IoT devices being produced dictates machine learning and deep learning approaches will be in the game as long as they exist.

Crypto-jacking - a form of Crypto-mining will continue to rise as there are more and more vulnerable devices to exploit.

Recent trends discussed in the introduction section, indicate that Ransomware attacks will continue to grow towards attacking small to medium scale business that is more vulnerable, and public organizations.

Malware family classification is anticipated to become more tedious, based on the studies [35], [96], that showcased the code reuse among IoT malware families to be having a rising trend.

The IoT malware is generally believed to be lacking sophistication compared to the Windows PE/Android malware counterparts [74], and most of the ML approaches ignore to tackle them under this assumption. Necessary precautions must be taken before malware authors bring in such sophistication as a surprise.

## VI. RELATED WORKS

There are notable Linux malware studies such as Cozzi *et al.* [28], [35]. In [28], they presented the first generic Linux-specific malware analysis pipeline, the first comprehensive Linux-based malware study covering low-level strategies employed by malware in the wild. And in [35], they showcased code reuse-based reconstruction of genealogy of IoT malware to aid in tracking their evolution over the years. Such lineage graphs are useful for learning the intra-family relationships among variants and aids in better family classification.

The IoTPOT (honeypot) based analysis work from Pa *et al.* [81] focuses on Telnet attacks targeting IoT devices, and the dataset they released publicly had been used by various studies such as [21], [22], [36], [40], [30], [78]. Some studies focus on a specific IoT malware family for dissection and analysis [28], such as Antonakakis *et al.* [97], where they performed analysis on the advent of the Mirai botnet over seven months period, the evolution of variants of Mirai, and the DDoS affected victim devices.

Wang *et al.* [98] studied Mirai to investigate brute-force attacks on IoT devices, analyze Darlloz and BASH-LITE to investigate exploitation of zero-day vulnerabilities in IoT devices. Forensic analysis of devices infected by Mirai botnet is provided by Zhang *et al.* [99]. Other botnet studies include analysis on Chuck Norris botnet by Čeleda *et al.* [100], [101], Dofloo/Spike botnet by Bohio [102], Psyb0t analysis by Durfina *et al.* [103], and Baume [104]. IoT *security-specific* survey was presented by HaddadPajouh *et al.* [105], intrusion detection system specific IoT review was provided by Khraisat and Alazab [106], DDoS attack mitigating intrusion detection systems are surveyed by Mishra and Pandya in [107].

We present the existing survey works by discussing general Linux surveys and IoT surveys first, followed by Cross-architecture specific IoT surveys.

### A. SURVEYS ON IoT MALWARE THREAT HUNTING

Aslan and Samet [108] provided a detailed analysis of general malware detection techniques, their uses, and drawbacks. The accounted malware detection approaches include those that deal with heuristics, behaviors, signatures, model-checking, deep learning, cloud, mobile, and IoT-based approaches.

Caviglione *et al.* [109] aimed at providing a bird's eye view of the development trends such as bio-inspired learning, transfer learning, and federated learning, and issues covering sophisticated techniques like information hiding and file-less malware. They also provided a meta-review over the existing malware detection-based surveys. They stated that there exists a lack of surveys in the literature over general detection frameworks that employ independent indicators for tackling heterogeneous scenarios. Our survey is, in fact, bridging such a gap for the IoT domain by surveying works on cross-architectural IoT malware threat hunting.

Costin and Zaddach [23] presented the first comprehensive analysis and survey on IoT malware, where they analyzed and produced reports about vulnerabilities, exploits, and defensive rules, for around 60 IoT malware families they collected. However, their work is inclined towards dynamic analysis of IoT malwares, for which they have also proposed an open-source analysis framework. At the time of this writing, the raw data and the analysis framework they released as part of their work [23] were not available.

Karanja et al. [110] focused on surveying general IoT malware characteristics. They elicit the need for IoT benchmark datasets and software tools for completeness in feature extraction attempts. However, their survey does not focus on cross-architecture IoT malware threat hunting; rather, it was discussed as research in need.

Milosevic et al. [111] covered software security incidents concerning IoT devices and discussed side-channel attacks such as differential power and fault analysis. Vignau et al. [96] discussed the feature-based evolution of IoT malware via multi-graph representation on feature propagation and also provided a taxonomy based on 16 behavioral characteristics of prominent IoT malwares. They also examined and found that reuse of features (malicious functionalities) is a more prevalent phenomenon among IoT malwares.

### B. SURVEYS ON CROSS-ARCHITECTURAL IoT MALWARE THREAT HUNTING

The problem of developing cross-architectural IoT malware threat hunting approaches is of recent interest to the research community, and there are only a very few works proposed due to practical issues as discussed in Section V. These factors also lead to scarcity of cross-architecture surveys. To the best of our knowledge, our survey is the first work dedicated specifically to the problem of cross-architectural IoT malware threat hunting.

The survey work by Ngo et al. [22] is the closest work, where the main theme of their survey was to cover static features-based ML approaches for IoT malware detection, including only three cross-architectural works, whereas we cover more than fifteen recent cross-architectural works. They employed a dataset with 7K malwares and 4K benign files to compare the approaches studied. Despite their attempt to evaluate studies, they did not provide architecture-wise reports such as architecture-wise distribution of samples in their dataset and corresponding model performances, making it a general Linux malware survey.

Moreover, the taxonomy they provided over existing static analysis approaches categorized based on feature types is too simplistic and failed to cover a wide range of valuable feature types. Their taxonomy covers only the following feature types, namely, ELF header [43], grayscale representations [78], strings [17], [21], opcodes [38], [112] [77], [113], opcode graphs [57], control flow graphs [41] and printable string graphs (PSI) [30].

### VII. CONCLUSION

An exponentially growing number of IoT devices had become a big data problem and a real challenge for machine learning-based threat hunting. We discussed the need for creating an OS platform-independent and CPU instruction set architecture-neutral threat hunting approaches and provided a modern taxonomy adopting the latest developments in the field of static analysis-based IoT malware threat hunting.

We then provided a comprehensive analysis of the existing cross-architectural research works and highlighted the challenges and opportunities open for future research.

At present, printable string information (PSI) is deemed to be the most useful cross-architectural feature followed by System and API calls. Graph-based methods are also showing strong potential but a long way to go to meet the constrained resource requirements of the IoT landscape.

With the recent release of the extended IoT malware binary dataset [87] from the IoTPOT team, a detailed evaluation of the cross-architectural approaches discussed in our surveyed research works using their dataset as benchmark is of much importance, and we plan to carry it out as a future work.

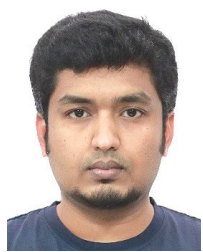

**ANANDHARAJU DURAI RAJU** received the B.Eng. degree in computer science from Anna University, Chennai, India. He is currently pursuing the Ph.D. degree with Simon Fraser University, Burnaby, Canada. He works as a Researcher (Intern) at Huawei Technologies Canada Company Ltd., Vancouver. He has more than seven years of previous professional work experience. His last role was a Technology Lead at Infosys Ltd. (Artificial Intelligence and Automation Services Unit). His research interests include data mining, machine learning, cybersecurity, and windows/linux malware. His current research is on the Internet of Things (IoT) malware.

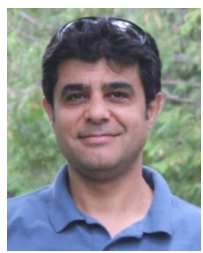

**IBRAHIM Y. ABUALHAOL** (Senior Member, IEEE) received the B.Sc., M.Sc., and Ph.D. degrees in electrical and computer engineering and the M.Eng. degree in technology innovation management. He is currently a Principal Data Scientist with Huawei Technologies and an Adjunct Research Professor with Carleton University, Ottawa, Canada. He is a Professional Engineer (P.Eng.) in ON, Canada. His research interests include machine learning and real-time big-data analytics and its applications in the Internet of Things, cybersecurity, and wireless communications.

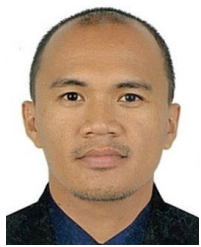

**RONNIE SALVADOR GIAGONE** received the B.S. degree in electronics and communications engineering from the Polytechnic University of the Philippines, Philippines, in 2003. In 2019, he joined the Huawei Technologies Group Company Ltd., as a Principal Security Researcher. His research interests include malware research, machine learning, and reverse engineering. In recent years, he has focused on researching new insights, theories, analyses, data, algorithms, and prototypes that advance cybersecurity technologies.

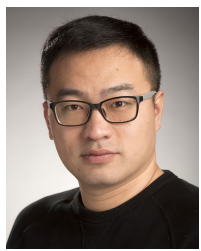

**YANG ZHOU** received the B.S. degree in electronic engineering from Jilin University, Changchun, China, in 2011, the M.S. degree in electronic engineering from Peking University, Beijing, in 2014, and the M.S. degree in computer science from Simon Fraser University, Burnaby, Canada, in 2017. He is currently a Data Scientist and a Cyber Security Researcher. His main research interests include malware detection and classification with machine learning techniques and attacks against operation systems and networks.

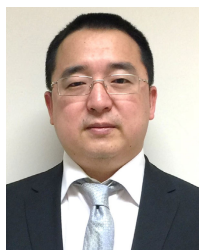

**SHENGQIANG HUANG** is currently a Senior Principal Engineer with Huawei Technologies Canada Company Ltd. He has more than 20 years of extensive experience in the network security industry, reverse engineering, video processing, financial trading system, and general software architecture with roles that span across organization and business maturity. His specialties are anti-virus engine development, intrusion prevention systems, sandboxing, industrial control systems security, APT/malware, NGFW, application layer security, and UTM. He is leading the anti-virus and sandbox product research in Huawei. Prior to joining Huawei, he was the Lead Architect and held the role of key research and development at Wurldtech Securities, a Canadian industrial control systems security startup acquired by General Electric (GE), in 2014. Prior to Wurldtech, he was a Senior Software Developer of Fortinet anti-virus engine core team. Before Fortinet, he had held various product management and development roles.

• • •